\documentclass[english,aps,preprint,showpacs]{revtex4}
\usepackage[T1]{fontenc}
\usepackage[latin9]{inputenc}
\usepackage{amssymb}
\usepackage{esint}
\usepackage{graphicx}

\makeatletter

\@ifundefined{textcolor}{}
{%
 \definecolor{BLACK}{gray}{0}
 \definecolor{WHITE}{gray}{1}
 \definecolor{RED}{rgb}{1,0,0}
 \definecolor{GREEN}{rgb}{0,1,0}
 \definecolor{BLUE}{rgb}{0,0,1}
 \definecolor{CYAN}{cmyk}{1,0,0,0}
 \definecolor{MAGENTA}{cmyk}{0,1,0,0}
 \definecolor{YELLOW}{cmyk}{0,0,1,0}
}

\makeatother

\usepackage{babel}
\begin{document}

\title{Bloch-Siegert shift of the Rabi model}

\author{Yiying Yan\footnote{yiyingyan@sjtu.edu.cn}}
\author{Zhiguo L\"{u}\footnote{zglv@sjtu.edu.cn}}
\author{Hang Zheng\footnote{hzheng@sjtu.edu.cn}}

\affiliation{Key Laboratory of Artificial Structures and Quantum Control (Ministry of Education), Department of Physics and Astronomy,Shanghai Jiao Tong University, Shanghai 200240, China}
\affiliation{Collaborative Innovation Center of Advanced Microstructures, Nanjing University, Nanjing 210093, China}
\date{\today}

\begin{abstract}
We apply a simple analytical method based on a unitary transformation to calculate the Bloch-Siegert (BS) shift over the entire driving-strength range. In quantitative comparison with the numerically exact BS shift obtained by Floquet formalism as well as the previous BS results, we confirm that our calculated results are not only accurate in the weak-driving regime but also correct in strong-driving limit. In the intermediate strong-driving regime, the calculated values of the BS shift are nearly the same as the exact ones. It turns out that our calculation for the BS shift is beyond perturbation. Meanwhile, we demonstrate the signatures caused by the BS shift by monitoring the excited-state population and the probe-pump spectrum under the experiment accessible conditions. In particular, we find that when the driving frequency is fixed at the transition frequency of the system, the lineshape of the probe-pump spectrum becomes asymmetric with the increase of the driving strength, which may be verified experimentally.
\end{abstract}
\maketitle

\section{introduction}
The physics of driven quantum systems, as an attractive topic in quantum physics, has been widely studied for several decades~\cite{Bloch,Shirley,Grifoni,Ashhab,Zheng1,Satanin}. At present, the study of such systems is significant for quantum information processing~\cite{Xu,Oliver,Economou,Greilich,Wiseman,Barnes} and renewed in the context of artificial atoms such as superconducting circuits~\cite{Nori,Yoshihara,Forn,Tuorila}. The prototype of driven quantum systems is the Rabi model describing a two-level system (TLS) driven by a harmonic driving (we set $\hbar=1$),
\begin{eqnarray}
  H(t)&=&\frac{1}{2}\omega_0\sigma_z+\frac{A}{2}\cos(\omega t)\sigma_x\nonumber\\
&=& \frac{1}{2}\omega_0\sigma_z+\frac{A}{4}(e^{i\omega t}\sigma_{-}+e^{-i\omega t}\sigma_{+})\nonumber\\
& &+\frac{A}{4}(e^{-i\omega t}\sigma_{-}+e^{i\omega t}\sigma_{+}),
\end{eqnarray}
where $\sigma_{x,y,z}$ is the usual Pauli matrix and $\sigma_\pm=(\sigma_x\pm i\sigma_y)/2$. $\omega_0$ is the transition frequency of the TLS. $A$ and $\omega$ are the amplitude and frequency of the harmonic driving, respectively.
Usually, one invokes the rotating-wave approximation (RWA), i.e. the omission of the counter-rotating (CR) terms $\frac{A}{4}(e^{-i\omega t}\sigma_{-}+e^{i\omega t}\sigma_{+})$~\cite{scully}. The RWA is valid in the weak driving and resonance cases, but breaks down in the strong driving regimes~[2]. It has been recognized that the CR terms lead to the shift of resonance position and additional beats in the time evolution~\cite{Bloch,Shirley}. Particularly, the shift of resonance position from resonance frequency of the RWA case is called the Bloch-Siegert (BS) shift: $\delta\omega_{\mathrm{BS}}$~\cite{Bloch,Shirley}. We focus on the BS shift in the present work.

The BS shift has been considered extensively in the cases of both classical and quantized fields~\cite{Bloch,Shirley,Stenholm,Hannaford,Cohen}. For the classical field, it is convenient to give the BS shift by the Floquet formalism developed by Shirley~\cite{Shirley}. For the quantized field, a fully quantum mechanical description was provided in the so-called dressed-atom model that combines the atom and driving field~\cite{Cohen}. All in all, most of pervious works, which are based on perturbation calculation, give the well-known BS shift in $6$th order of $A$~\cite{Shirley}. Hence, there is not yet a simple analytical method to produce the BS shift in the entire range of the driving strength.

Apart from the theoretical calculation of the BS shift, the effect of the BS shift has been studied before, such as an asymmetric Autler-Townes profile in a driven three-level system~\cite{Wei}, the shift of the sidebands of resonance fluorescence~\cite{Browne,Zheng2}, application to the astrophysical determination of the fundamental constants variation~\cite{Solovyev}, etc. More recently, we showed that the BS-type correction plays an important role in the relaxation and dephasing processes~\cite{Zheng3}. In addition, the BS shift has also been experimentally observed in strongly driven artificial atoms~\cite{Forn,Tuorila}.

In this paper, we calculate the BS shift over the entire driving-strength regime ($0<A/\omega_0<\infty$) by a simple analytical method based on a unitary transformation~\cite{Zheng1}. The approach has been used to study the driven tunneling dynamics of the TLS in our previous work~\cite{Zheng1}. It turned out that our method provides almost the same description as the numerically exact treatment over a wide parameter regime of interest. Moreover, we prove that the BS shift up to the $4$th order in $A$ given by our method is the same as that obtained by the Floquet approach~\cite{Shirley}. In this work, we show that both small and large BS shifts can be uniformly and correctly evaluated from the derivative of effective Rabi frequency we derived, in comparison with numerically exact results and previous analytical results. Moreover, we illustrate that the signatures of the BS shift can be monitored in the emission process and probe-pump spectrum under the experiment accessible conditions. When $\omega=\omega_0$, we show that the sidebands of the non-RWA probe-pump spectrum are generally asymmetric. The asymmetry results from the BS shift and can be enhanced with increasing the driving strength. On the other hand, the non-RWA probe-pump spectra become symmetric only when $\omega=\omega_0+\delta\omega_{\rm BS}$. According to these properties, one may check the signatures of the BS shift experimentally.

\section{Unitary transformation\label{sec:Unitary-transformation}}

The wave function $|\Psi(t)\rangle$ of the TLS satisfies Schr\"{o}dinger equation
\begin{equation}
  \left[H(t)-i\frac{d}{dt}\right]|\Psi(t)\rangle=0.
\end{equation}
We transform the above equation as follows:
\begin{equation}
  e^{S(t)}\left[H(t)-i\frac{d}{dt}\right][e^{-S(t)}e^{S(t)}|\Psi(t)\rangle]\equiv \left[H^\prime(t)-i\frac{d}{dt}\right]|\Psi^\prime(t)\rangle=0,
\end{equation}
where $|\Psi^\prime(t)\rangle=e^{S(t)}|\Psi(t)\rangle$ and
\begin{equation}
  H^{\prime}(t)=e^{S(t)}H(t)e^{-S(t)}-ie^{S(t)}\frac{d}{dt}e^{-S(t)}.
\end{equation}
Here,  we give the time-dependent generator $S(t)$ of the transformation
\begin{equation}
  S(t)=i\frac{A}{2\omega}\xi\sin(\omega t)\sigma_{x},
\end{equation}
with parameter $\xi\in [0,1]$ to be determined later~\cite{Zheng1}. Thus, we obtain the transformed Hamiltonian
\begin{eqnarray}
H^{\prime}(t) & = & \frac{1}{2}\omega_{0}\left\{ \cos\left[\frac{A}{\omega}\xi\sin(\omega t)\right]\sigma_{z}+\sin\left[\frac{A}{\omega}\xi\sin(\omega t)\right]\sigma_{y}\right\} \nonumber \\
 &  & +\frac{A}{2}(1-\xi)\cos(\omega t)\sigma_{x}.
\end{eqnarray}
We make use of the identity
$
\exp\left[i\frac{A}{\omega}\xi\sin(\omega t)\right]=\sum_{n=-\infty}^{\infty}J_{n}\left(\frac{A}{\omega}\xi\right)\exp(in\omega t),
$
in which $J_{n}(\cdot)$ is the $n$th-order Bessel function
of the first kind, and divide the Hamiltonian into three parts $H^{\prime}(t)=H_{0}^{\prime}+H_{1}^{\prime}(t)+H_{2}^{\prime}(t)$,
\begin{eqnarray}
H_{0}^{\prime} & = & \frac{1}{2}\omega_{0}J_{0}\left(\frac{A}{\omega}\xi\right)\sigma_{z}, \\
H_{1}^{\prime}(t) & = & \frac{A}{2}(1-\xi)\cos(\omega t)\sigma_{x}+\omega_{0}J_{1}\left(\frac{A}{\omega}\xi\right)\sin(\omega t)\sigma_{y},\\
H_{2}^{\prime}(t) & = & \omega_{0}\sum_{n=1}^{\infty}J_{2n}\left(\frac{A}{\omega}\xi\right)\cos(2n\omega t)\sigma_{z}\nonumber \\
 &  & +\omega_{0}\sum_{n=1}^{\infty}J_{2n+1}\left(\frac{A}{\omega}\xi\right)\sin[(2n+1)\omega t]\sigma_{y},
\end{eqnarray}
where $H_{2}^{\prime}(t)$ includes all higher-order harmonic terms ($n\geq 2$). Up till now, the treatment is exact. To proceed, we neglect the part $H_2^\prime(t)$ since the higher-order harmonic terms with the higher-order Bessel functions are reasonably negligible over a wide range of parameters space~\cite{Zheng1}. Therefore, we arrive at the Hamiltonian $H^{\prime}(t)\simeq H_{0}^{\prime}+H_{1}^{\prime}(t)$. We verify that this treatment is beyond perturbation by the correct prediction of the BS shift for $A/\omega_0\in[0,\infty)$.

To proceed, we determine $\xi$ by
\begin{equation}
J_1\left(\frac{A}{\omega}\xi\right)\omega_{0}=\frac{A}{2}(1-\xi)\equiv\frac{\tilde{A}}{4}.\label{eq:xieq}
\end{equation}

Consequently, we rewrite the Hamiltonian $H^{\prime}(t)$ as
\begin{equation}\label{Hprime}
H^{\prime}(t)= \frac{1}{2}J_{0}\left(\frac{A}{\omega}\xi\right)\omega_{0}\sigma_{z}+\frac{\tilde{A}}{4}(e^{-i\omega t}\sigma_{+}+e^{i\omega t}\sigma_{-}).
\end{equation}
This is the counter-rotating hybridized rotating wave (CHRW) Hamiltonian~\cite{Zheng1}. Note that the CHRW Hamiltonian possesses a RWA-like form with a renormalized transition frequency $J_{0}\left(\frac{A}{\omega}\xi\right)\omega_{0}$ and a renormalized driving strength $\tilde{A}$. The renormalized quantities in the transformed Hamiltonian results explicitly from the effects of CR interactions \cite{Zheng1}.

We apply a rotating operation $R(t)=\exp\left(\frac{i}{2}\omega t\sigma_{z}\right)$ to the Hamiltonian (\ref{Hprime}), and get
\begin{equation}
\tilde{H}=\frac{\tilde{\Delta}}{2}\sigma_{z}+\frac{\tilde{A}}{4}\sigma_{x},
\end{equation}
where $\tilde{\Delta}=J_{0}\left(\frac{A}{\omega}\xi\right)\omega_{0}-\omega$ is the effective detuning. The Hamiltonian $\tilde{H}$ can be readily diagonalized. Its eigenenergies are
\begin{equation}
  \varepsilon_{\pm}=\pm\frac{1}{2}\tilde{\Omega}_R,
\end{equation}
with $\tilde{\Omega}_R=\sqrt{\tilde{\Delta}^{2}+\tilde{A}^{2}/4}$ being the effective Rabi frequency. The corresponding eigenstates, i.e. dressed states are
\begin{equation}
  |\widetilde{\pm}\rangle=\sin\theta|\mp\rangle\pm\cos\theta|\pm\rangle,
\end{equation}
with $\theta=\arctan\left[(\tilde{\Omega}_{R}-\tilde{\Delta})/(\tilde{A}/2)\right]$ and $|\pm\rangle$ being the bare bases: $\sigma_z|\pm\rangle=\pm|\pm\rangle$.

%We show the relation between $\tilde{\Omega}_R$ in our method and quasienergy in the Floquet formalism~\cite{Shirley} in order to understand the role of  $\tilde{\Omega}_R$ on the evaluation of the BS shift in the next section. Taking $|\Psi_\pm(0)\rangle=|\widetilde{\pm}\rangle$, we find that the states of TLS evolve as
%\begin{equation}
%  |\Psi_\pm(t)\rangle=e^{-S(t)}R^\dagger(t)e^{-i\tilde{H}t}|\widetilde{\pm}\rangle =e^{-i\frac{1}{2}(\omega\pm\tilde{\Omega}_R)t}e^{i\frac{1}{2}\omega t-S(t)}R^\dagger(t)|\widetilde{\pm}\rangle\equiv e^{-i\frac{1}{2}(\omega\pm\tilde{\Omega}_R)t} |\phi_\pm(t)\rangle.
%\end{equation}
%According to Floquet theorem, we immediately verify that $\frac{1}{2}(\omega\pm\tilde{\Omega}_R)$ play the same role as quasienergies and $|\phi_\pm(t)\rangle=|\phi_\pm(t+2\pi/\omega)\rangle$ are the so-called Floquet modes~\cite{Shirley,Grifoni}.

\section{Bloch-Siegert Shift}

\subsection{Evaluation of the Bloch-Siegert shift}

We calculate the BS shift by using the effective Rabi frequency obtained in the former section.
The BS shift measures the deviation of resonance frequency $\omega_{\mathrm{res}}$ of the original Hamiltonian from that of the RWA:
\begin{equation}\label{bs}
\delta\omega_{\mathrm{BS}}=\omega_{\mathrm{res}}-\omega_{0},
\end{equation}
where $\omega_{0}$ is the resonance frequency of the RWA case.
In general, the resonance frequency $\omega_{\mathrm{res}}$ can be determined from the time-averaged transition probability in Ref.~\cite{Shirley},
$
\overline{P}=\left[1-4\left(\partial q_\alpha/\partial\omega_{0}\right)^{2}\right]/2,
$
where $q_\alpha$ is the quasienergy of the Floquet Hamiltonian. In principal, on solving the equation $\partial q_\alpha/\partial\omega_{0}=0$ for driving frequency $\omega$, we determine the position of the resonant transitions. Alternatively, we can also use $\partial q_\alpha^2/\partial\omega_{0}=0$ as the resonance condition in the analytical calculation instead of $\partial q_\alpha/\partial\omega_{0}=0$.

Similar to Shirley's method, the resonance condition in our formalism is $\partial\tilde{\Omega}_{R}^{2}/\partial\omega_{0}=0$ since $q_\alpha$ is related to $\tilde{\Omega}_{R}$ by the relation $q_\pm=(\omega\pm\tilde{\Omega}_R)/2$. By solving the equation for variable $\omega$, we get the resonance frequency $\omega_{\mathrm {res}}$. We can readily derive the explicit form for the derivative $\partial\tilde{\Omega}_{R}^{2}/\partial\omega_{0}$, which
reads
\begin{eqnarray}
\frac{\partial\tilde{\Omega}_{R}^{2}}{\partial\omega_{0}} & = & 2\left[\omega_{0}J_{0}\left(\frac{A}{\omega}\xi\right)-\omega\right]\left[J_{0}\left(\frac{A}{\omega}\xi\right)-\omega_{0}\frac{A}{\omega}\right.\nonumber\\
 &  & \left.\times J_{1}\left(\frac{A}{\omega}\xi\right)\frac{\partial\xi}{\partial\omega_{0}}\right]-2A^{2}(1-\xi)\frac{\partial\xi}{\partial\omega_{0}},\label{eq:drabi}
\end{eqnarray}
where $\partial\xi/\partial\omega_0$ is determined from Eq.~(\ref{eq:xieq}) and takes the form
\begin{equation}
\frac{\partial\xi}{\partial\omega_{0}}=-\frac{2\omega J_{1}\left(\frac{A}{\omega}\xi\right)}{A\left\{ \omega+\omega_{0}\left[J_{0}\left(\frac{A}{\omega}\xi\right)-J_{2}\left(\frac{A}{\omega}\xi\right)\right]\right\}}.\label{eq:dxi}
\end{equation}
Using Eqs.~(\ref{eq:xieq}),~(\ref{eq:drabi}) and~(\ref{eq:dxi}), we can self-consistently determine $\xi$ and $\omega$. The obtained $\omega$ is actually the desired resonance frequency $\omega_{\mathrm{res}}$ of the Rabi Hamiltonian. Therefore,
it is straightforward to obtain the BS shift by Eq.~(\ref{bs}). On one hand, in the weak driving case, we have correctly given the power series expansion of the BS shift in the previous work~\cite{Zheng1}. On the other hand, it is easy to solve a large BS shift in the strong-driving limit. When $A/\omega_0\rightarrow\infty$, $\xi\rightarrow 1$ and $\partial\xi/\partial\omega_0\rightarrow0$. Thus, the solution to Eq.~(\ref{eq:drabi}) for $\omega$ is determined by $ J_{0}\left(\frac{A}{\omega}\xi\right)=0$. Therefore, the first zero of $J_0\left(\frac{A}{\omega}\xi\right)$ gives the resonance frequency in the strong-driving limit,
\begin{equation}
\omega_{\mathrm{res}}=A/2.404826. \label{eq:asym}
\end{equation}
This result is the same as that obtained in Ref.~\cite{Shirley}.

An interesting phenomena, coherent destruction of tunneling (CDT), occurs at $A/\omega=2.404826$ ~\cite{Zheng1,Thorwart}. Surprisingly, we immediately recognize that the CDT condition coincides with the resonance condition Eq.(\ref{eq:asym}). Hence, it means that the CDT is actually a resonantly-driven phenomenon. In other words, it indicates that the large BS shift plays important role in this unique phenomenon.

\subsection{Numerical validation and comparison}
In order to illustrate the validity of our method and the accuracy of the calculated BS shift, we compare our results with the numerically exact and previous analytical results given by the methods based on Floquet theory. We show the approach of calculating the exact numerical BS shift by Floquet formalism. As shown in Ref.~\cite{Shirley}, the derivative $\frac{\partial q_\alpha}{\partial\omega_{0}}$ can be evaluated from the eigenstates of the Floquet Hamiltonian
\begin{equation} \label{shirleyqw}
\frac{\partial q_{\alpha}}{\partial\omega_{0}}=\sum_{\gamma l}a_{\gamma}|\langle\gamma l|\lambda_{\alpha0}\rangle|^{2},
\end{equation}
where the notations are the same as Shirley's (the index $\gamma$ represents states of the TLS and $l$ denotes integer representing Fourier component. $|\lambda_{\alpha0}\rangle$ is the eigenstate of Floquet Hamiltonian associated with quasienergy $q_{\alpha}$. The coefficient $a_{\gamma}=-(+)\frac{1}{2}$
if $\gamma$ represents the ground state (excited state)). For a fixed $A$ and given $\omega$, we first numerically diagonalize the Floquet Hamiltonian (9) in Ref.~\cite{Shirley} with an appropriate truncation, and then get the eigenstate $|\lambda_{\alpha0}\rangle$. Using Eq.~(\ref{shirleyqw})
we can numerically obtain the value of derivative $\frac{\partial q_{\alpha}}{\partial\omega_{0}}$
with the fixed $A$ and $\omega$. It is therefore possible to determine the resonance frequency by numerically searching the solution to the equation $\frac{\partial q_{\alpha}}{\partial\omega_{0}}=0$
or the minimum position of $\left(\frac{\partial q}{\partial\omega_{0}}\right)^{2}$. The BS shift obtained in this way is referred to numerically exact result.

We introduce two approximate methods for comparison by the analytical result of the $q$ obtained by Shirley~\cite{Shirley,Hannaford}. First, in terms of the analytical result $\frac{\partial q ^2}{\partial\omega_{0}}=0$, we obtain
\begin{eqnarray}
\omega & = & \omega_{0}+\frac{\omega A^{2}}{4(\omega+\omega_{0})^{2}}+\frac{(2\omega_{0}-\omega)A^{4}}{64(\omega+\omega_{0})^{4}}\nonumber \\
 &  & +\frac{(9\omega^{5}-126\omega^{4}\omega_{0}+82\omega^{3}\omega_{0}^{2}+42\omega^{2}\omega_{0}^{3}-23\omega\omega_{0}^{4}-8\omega_{0}^{5})A^{6}}{256(\omega+\omega_{0})^{6}(9\omega^{2}-\omega_{0}^{2})^{2}},\label{q2}
\end{eqnarray}
%\begin{eqnarray}\label{q2}
%q^{2}&=&\frac{1}{4}(\omega-\omega_{0})^{2}+\frac{2\omega_{0}(A/4)^{2}}{\omega+\omega_{0}}-\frac{2\omega_{0}(A/4)^{4}}{(\omega+\omega_{0})^{3}}\nonumber\\
%& &+\frac{8\omega_{0}(\omega^{2}-5\omega\omega_{0}-2\omega_{0}^{2})(A/4)^{6}}{(\omega+\omega_{0})^{5}(9\omega^{2}-\omega_{0}^{2})},\label{eq:squareq}
%\end{eqnarray}
and calculate the BS shift for comparison. On one hand, we can iteratively solve Eq.~(\ref{q2}) for the resonance frequency $\omega$ and gets the BS shift. It has been pointed out that this equation can be used to calculate the
whole range of the shift from 0 to $100\%$~\cite{Hannaford}. On the other hand, carrying out perturbation calculation, we can derive the BS shift up to 6th order in $A$, which reads~\cite{Shirley}
\begin{equation} \label{bs6th}
  \delta\omega_{\rm{BS}}^{(6\rm{th})}=\frac{(A/4)^2}{\omega_0}+\frac{(A/4)^4}{4\omega_0^3}-\frac{35(A/4)^6}{32\omega_0^{5}}.
\end{equation}
In what follows, we show the comparison among various results.

\begin{figure}
  % Requires
  \includegraphics[width=8cm]{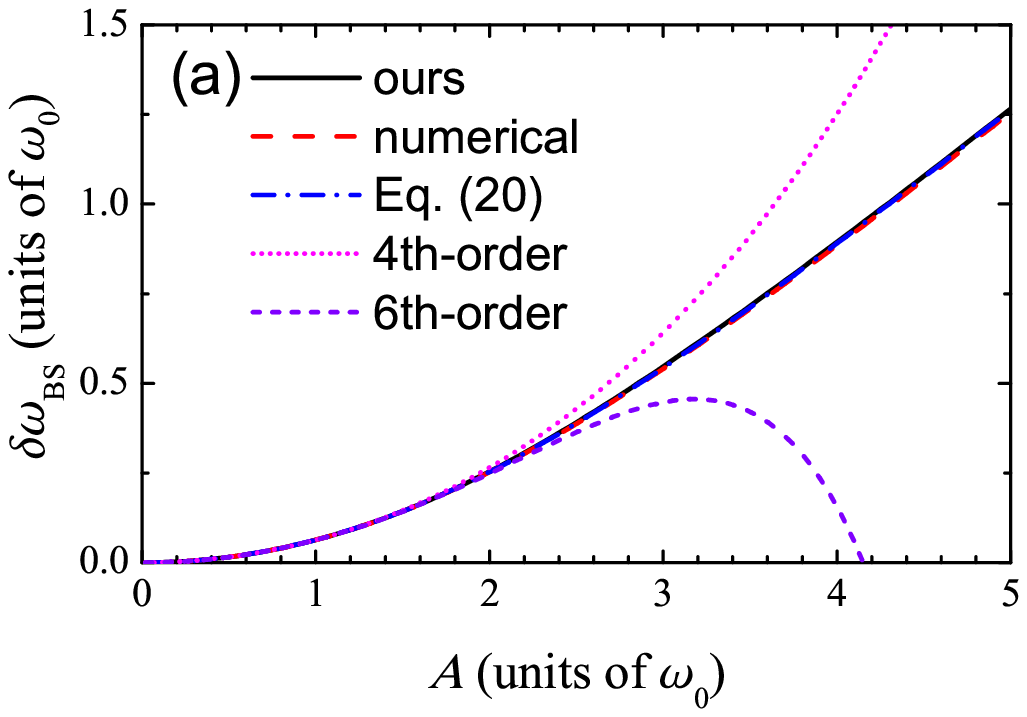}
  \includegraphics[width=8cm]{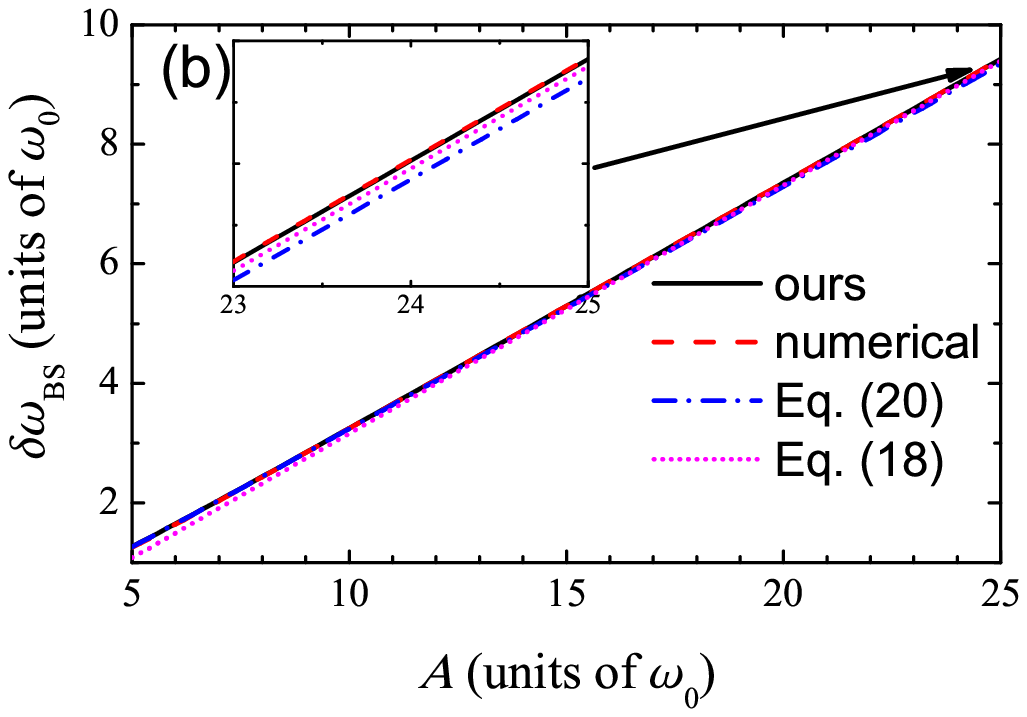}\\
  \caption{(Color online) The BS shift as a function of driving strength $A$ for various methods. The precision of numerical method is defined by the deviation of $\overline{P}$ from the maximum $1/2$, which is about $10^{-15}$ for given $A$ and $\omega_\mathrm{res}$ in this work.}\label{fig1}
\end{figure}

In Figs.~\ref{fig1}(a)-\ref{fig1}(b), we make a quantitative comparison between the values of the BS shift obtained from our method and those given by the other approaches mentioned above over a wide driving strength range. From weak-driving to moderately strong-driving regime, we notice that our result is in good agreement with the numerically exact result and that of Eq.(\ref{q2}). At the same time, it is obvious to see that the results of both 4th-order and 6th-order BS shifts differ from those of the other methods when $A/\omega_{0}>2$. It means that the BS shift up to 6th order of $A$ is not adequate for moderately strong driving case and higher-order terms need to be considered when $A$ increases. In addition, in Fig.~\ref{fig1}(b), we find that our result is almost the same as the numerically exact result when $A/\omega_{0}>5$ but slightly differs from Eq.~(\ref{q2}) when the driving is sufficiently strong [see the inset in Fig.~\ref{fig1}(b)]. Moreover, it is obvious to see that the results given by Eq.~(\ref{eq:asym}) provides a better description of the position of the resonance than that of Eq.~(\ref{q2}) because the result given by Eq.~(\ref{eq:asym}) becomes more and more accurate as $A\rightarrow\infty$.

\begin{figure}
  \includegraphics[width=8cm]{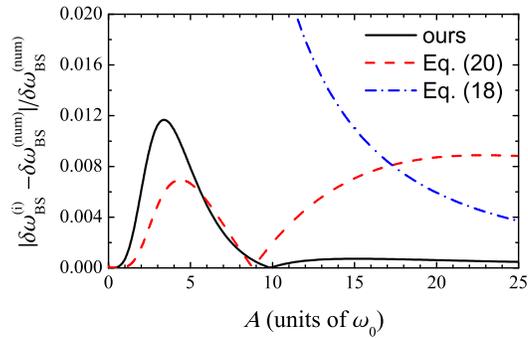}\\
  \caption{(Color online) The deviation between analytical results and numerical result.}\label{fig2}
\end{figure}

In order to show the accuracy of our method in detail, we illustrate the quantitative difference of the BS shift among analytical and numerical
approaches. First, we show the values of the BS shift calculated by the four methods in Table~\ref{tab:table1} for comparison. Second, in Fig.\ref{fig2}, we show the deviation $|\delta\omega_{{\rm BS}}^{(i)}-\delta\omega_{{\rm BS}}^{({\rm num})}|/\delta\omega_{{\rm BS}}^{({\rm num})}$,
in which $\delta\omega_{{\rm BS}}^{({\rm num})}$ denotes the numerically
exact BS shift and $\delta\omega_{{\rm BS}}^{(i)}$ represents the
BS shift obtained by one of the three analytical methods, i.e., our
method, Eqs.~(\ref{q2}) and~(\ref{eq:asym}).
We find that when $2.5<A/\omega_{0}<4.5$, the deviation of our result from the numerically exact result is less than $1.2\%$, whereas the deviation of Eq.~(\ref{q2}) is less than $1\%$. Nevertheless, when $A<2.5\omega_0$ or $A>4.5\omega_0$, the deviation of our result is less than $1\%$. In particular, when $A/\omega_{0}>9$, our method becomes the most accurate among the three analytical methods,
with the deviation being less than $0.02\%$. In addition, Eq.~(\ref{eq:asym}) becomes more accurate than the Eq.~(\ref{q2}) only for
$A/\omega_{0}>17$.

To sum up, the BS shift given by our method is in quantitatively good agreement with the numerical result over the entire range $0<A/\omega_0<\infty$. Thus, it turns out that our calculation for the BS shift is beyond perturbation and the higher-order corrections involved in the BS shift have been correctly taken into account in our method. Moreover, our method is applicable to solve the dynamics of the TLS in the presence of the dissipation. As a result, we could demonstrate the measurable signatures of the BS shift in the excited-state population and the probe-pump spectrum.

\section{signatures of the Bloch-Siegert shift}
In general, the signature of the BS shift can be measured in either the emission or absorption process. In order to correctly discuss the observable signatures indicating the resonance, we should take into account the dissipation induced by the coupling to the environment in the Rabi model.
We consider that the radiation of the TLS is described
by the following master equation
\begin{eqnarray}
\frac{d}{dt}\rho(t)&=&-i[H(t),\rho(t)]-\frac{\kappa}{2}[\sigma_{+}\sigma_{-}\rho(t)\nonumber\\& &+\rho(t)\sigma_{+}\sigma_{-}-2\sigma_{-}\rho(t)\sigma_{+}],\label{eq:me}
\end{eqnarray}
where $\rho(t)$ is the reduced density matrix of the TLS and $\kappa$ is the decay rate. The solution of the master equation has been obtained as a continued fraction in Ref.~\cite{Stenholm}. Here, we apply the unitary transformation to the master equation so as to transform it into a feasible form.

By employing the relation between the transformed density matrix $\tilde{\rho}(t)$ and the original density matrix $\rho(t)$: \[\tilde{\rho}(t)=R(t)e^{S(t)}\rho(t)e^{-S(t)}R^\dagger(t),\] we derive the equation of motion for the element in the basis of $|\widetilde{\pm}\rangle$, $\tilde{\rho}_{\alpha\beta}(t)=\langle\widetilde{\alpha}|\tilde{\rho}(t)|\widetilde{\beta}\rangle$:
\begin{equation}
\frac{d}{dt}\tilde{\rho}_{\alpha\beta}(t)=-i(\varepsilon_{\alpha}-\varepsilon_{\beta})\tilde{\rho}_{\alpha\beta}(t)-\sum_{\mu,\nu}\mathcal{L}_{\alpha\beta,\mu\nu}\tilde{\rho}_{\mu\nu}(t),\label{eq:menew}
\end{equation}
where a Greek index denotes $\pm$ and
\begin{eqnarray}
\mathcal{L}_{\alpha\beta,\mu\nu}&=&\frac{\kappa}{2}\sum_{n,\,\lambda}\left\{ \delta_{\nu,\beta}X_{\alpha\lambda,n}^{+}X_{\lambda\mu,-n}^{-}+\delta_{\mu,\alpha}X_{\nu\lambda,n}^{+}X_{\lambda\beta,-n}^{-}\right.\nonumber\\
& &\left.-2X_{\alpha\mu,n}^{-}X_{\nu\beta,-n}^{+}\right\},
\end{eqnarray}
with $n$ taking on integer.
The derivation of Eq.~(\ref{eq:menew}) is given in Appendix~\ref{sub:The-derivation} and the explicit expressions for $X_{\alpha\beta,k}^{\pm}$ can be found therein. Moreover, the solution to the equation used in the following are given in Appendix~\ref{app:sols}. Interestingly, we find that the derived master equation takes the similar structure as the Floquet-Born-Markov master equation given in Ref.~\cite{Grifoni2}, which is derived from the total Hamiltonian consisting of the TLS, the reservoir, and their interaction. In the following, we calculate time-averaged population of TLS and the probe-pump spectrum based on the time-independent master equation we derived so as to show the detectable signals caused by the BS shift.

\subsection{The population of excited state of TLS }
Since the emission process is related to the steady-state population of excited state of the TLS, we attempt to find the signature of the BS shift by monitoring the population \cite{Stenholm}. Moreover, the total intensity of resonance fluorescence is proportional to this population in the steady state~\cite{Mollow2}. In terms of the dressed-state population difference $\langle\tilde{s}_z\rangle_\mathrm{ss}=\tilde{\rho}_{++}(\infty)-\tilde{\rho}_{--}(\infty)$, we give the expression for the steady-state population of TLS by $\tilde{\rho}(t)$,
\begin{eqnarray}
  \rho_{++}(\infty)&=&\lim_{t\rightarrow\infty}\mathrm{Tr}[|+\rangle\langle+|\rho(t)]\nonumber\\
  &=&\lim_{t\rightarrow\infty}\mathrm{Tr}[R(t)e^{S(t)}|+\rangle\langle+|e^{-S(t)}R^\dagger(t)\tilde{\rho}(t)]\nonumber\\
  &=&\frac{1}{2}+\frac{1}{2}\langle\tilde{s}_z\rangle_{\mathrm{ss}}\left\{ \cos(2\theta)\cos\left[\frac{A}{\omega}\xi\sin(\omega t)\right]\right.\nonumber \\
 &  & \left.+\sin(2\theta)\sin(\omega t)\sin\left[\frac{A}{\omega}\xi\sin(\omega t)\right]\right\}\nonumber\\
 & = & \frac{1}{2}\left\{1+\langle\tilde{s}_z\rangle_{\mathrm{ss}}\left[\cos(2\theta)J_{0}\left(\frac{A}{\omega}\xi\right)+\sin(2\theta)J_{1}\left(\frac{A}{\omega}\xi\right)\right]\right\}\nonumber \\
 &  & +\frac{1}{2}\langle\tilde{s}_z\rangle_{\mathrm{ss}}\sum_{n=1}^{\infty}\left\{2\cos(2\theta)J_{2n}\left(\frac{A}{\omega}\xi\right)\right.\nonumber \\
 &  & \left.+\sin(2\theta)\left[J_{2n+1}\left(\frac{A}{\omega}\xi\right)-J_{2n-1}\left(\frac{A}{\omega}\xi\right)\right]\right\}\cos(2n\omega t).
\end{eqnarray}
When deriving the expression above, we have neglected the terms proportional to $\langle\tilde{s}_\pm\rangle_{\mathrm{ss}}$, which are of order $\kappa/\tilde{\Omega}_R$.
We find that the even multiple frequencies appear
resulting from the unitary transformation. This property of the
diagonal elements is similar as that of the ansatz used in Ref.~\cite{Stenholm}.

Recalling that normal measurement of the experiment takes time average, we concentrate on the interested quantity, the time-averaged
population $\overline{\rho_{++}}$ from $\rho_{++}(\infty)$, which reads
\begin{eqnarray}\label{eq:rhopp}
\overline{\rho_{++}}&=&\frac{1}{2}\left\{1+\left\langle \tilde{s}_{z}\right\rangle _{{\rm ss}}\left[\cos(2\theta)J_{0}\left(\frac{A}{\omega}\xi\right)+\sin(2\theta)J_{1}\left(\frac{A}{\omega}\xi\right)\right]\right\}\nonumber\\
&=&\frac{1}{2}-\frac{\gamma_0^2}{2\kappa\gamma_z},
\end{eqnarray}
where we used $\gamma_0\simeq\kappa\left[\cos(2\theta)J_{0}\left(\frac{A}{\omega}\xi\right)+\sin(2\theta)J_{1}\left(\frac{A}{\omega}\xi\right)\right]$ and $\langle\tilde{s}_z\rangle_\mathrm{ss}\simeq-\gamma_0/\gamma_z$.

\begin{figure}
  \includegraphics[width=8cm]{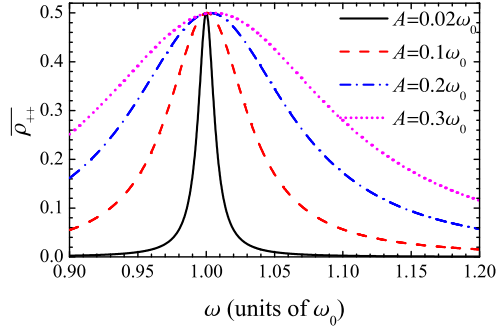}\\
  \caption{(Color online) The time-averaged population $\overline{\rho_{++}}$ as a function of driving frequency $\omega$ for $\kappa=2\times10^{-3}\omega_0$ and various driving strength $A$.}\label{fig3}
\end{figure}

In Fig.~\ref{fig3}, we show the time-averaged population $\overline{\rho_{++}}$
as a function of driving frequency $\omega$ for various driving strengths $A$. The results
show that the resonance is not only shifted but also broadened as $A$ increases. These results are consistent
with those of the previous work based on a continued fraction technique~\cite{Stenholm}. However, we point out that $\overline{\rho_{++}}$ is physically related to the dressed-state population difference in our formalism, which is more clear than that in Ref.~\cite{Stenholm}. More importantly, Eq.~(\ref{eq:rhopp}) indicates that the BS shift influences the dressed-state population difference. It means that probe-pump spectra may exhibit the effects of BS shift. In addition, according to these results, it is reasonable to predict that for the non-RWA case, the maximum intensity of resonance fluorescence can be achieved only when $\omega=\omega_0+\delta\omega_{\rm BS}$ for a fixed $A$.

\subsection{The probe-pump spectrum}
In this section, we show the signature of BS shift from the probe-pump
spectrum. It is well-known that the probe-pump spectrum is sensitive to the dressed-state population difference. A non-zero population difference of dressed states can cause the lineshape of the spectrum to be asymmetric, which has two unequal sidebands~\cite{Mollow,Tanas}. In consequence of this property of the spectrum, we can expect that the spectrum may be sensitive to the BS shift because the BS shift can modify the dressed-state population. According to the linear response theory, the steady-state response of driven TLS to a weak probe field
\[
H_{p}(t)=\Omega_{p}(\sigma_{+}e^{-i\upsilon t}+\sigma_{-}e^{i\upsilon t}),
\]
in which $\Omega_{p}$ and $\upsilon$ are the amplitude and frequency of
the probe field,
is given by the Fourier transform of the two-time commutator~\cite{Mollow}
\begin{equation}
S(\upsilon)\propto\mathrm{Re}\lim_{t\rightarrow\infty}\int_{0}^{\infty}\left\langle [\sigma_{-}(t+\tau),\sigma_{+}(t)]\right\rangle e^{i\upsilon\tau}d\tau,\label{eq:sfun}
\end{equation}
where $\sigma_{\pm}(t)\equiv U^{\dagger}(t)\sigma_{\pm}U(t)$ are the operators in the Heisenberg picture with $U(t)$ being the evolution
operator for the total system consisting of the TLS, driving field and reservoir causing the relaxation (in the absence of the probe field). We now apply our unitary transformation to the two-time commutator so that it may be related to the solution to Eq.~(\ref{eq:menew}).

We first consider how to evaluate the two-time correlation function $\left\langle \sigma_{-}(t+\tau)\sigma_{+}(t)\right\rangle$ by making use of the unitary transformation and rotation operation because the other function $\left\langle \sigma_{+}(t)\sigma_{-}(t+\tau)\right\rangle$ can be treated similarly. The two-time correlation function is evaluated in the original frame as follows~\cite{Mollow}:
\begin{eqnarray}
\left\langle \sigma_{-}(t+\tau)\sigma_{+}(t)\right\rangle  & = & \mathrm{Tr}_{SR}[U^{\dagger}(t+\tau)\sigma_{-}U(t+\tau)U^{\dagger}(t)\sigma_{+}U(t)\rho(0)\rho_{R}]\nonumber \\
 & = & \mathrm{Tr}_{SR}[U(t)U^{\dagger}(t+\tau)\sigma_{-}U(t+\tau)U^{\dagger}(t)\sigma_{+}\rho(t)\rho_{R}],
\end{eqnarray}
where $\rho(0)$ is the initial state of the TLS and $\rho_R$ is the state of the reservoir.
By employing
$\tilde{U}(\tau)=R(t+\tau)e^{S(t+\tau)}U(t+\tau)U^{\dagger}(t)e^{-S(t)}R^{\dagger}(t)$ (we assume that the evolution is $\tau$-dependent in the transformed frame) and with the aid of identities (\ref{eq:identity}), we have
\begin{eqnarray}
\left\langle \sigma_{-}(t+\tau)\sigma_{+}(t)\right\rangle  & = & \mathrm{Tr}_{SR}[\tilde{U}^{\dagger}(\tau)R(t+\tau)e^{S(t+\tau)}\sigma_{-}e^{-S(t+\tau)}R^{\dagger}(t+\tau)\nonumber \\
 &  & \times\tilde{U}(\tau)R(t)e^{S(t)}\sigma_{+}e^{-S(t)}R^{\dagger}(t)\tilde{\rho}(t)\rho_{R}]\nonumber \\
 & = & \frac{1}{4}\sum_{n,l\,{\rm odd}}\mathrm{Tr}_{SR}[\tilde{U}^{\dagger}(\tau)(\hat{C}_{n}^{\dagger}e^{-in\omega(t+\tau)}+\hat{D}_{n}^{\dagger}e^{in\omega(t+\tau)})\nonumber \\
 &  & \times\tilde{U}(\tau)(\hat{C}_{l}e^{il\omega t}+\hat{D}_{l}e^{-il\omega t})\tilde{\rho}(t)\rho_{R}],\label{eq:commutator}
\end{eqnarray}
where the summation is taken over all positive odd integers, and $\hat{C}_n$ and $\hat{D}_n$ are defined in Eqs.~(\ref{eq:cn}) and (\ref{eq:dn}).

To proceed, we take the long-time limit ($t\rightarrow\infty$) and neglect the $t$-dependent
terms in Eq.~(\ref{eq:commutator}) because their contributions are negligible to a long-time observation.
Besides, one can verify that the terms with the factor $e^{in\omega\tau}$
contribute a finite value to the spectrum only when $\upsilon<0$. It
is therefore reasonable to omit these terms when one considers the
probe-pump spectrum. Consequently, we arrive at the following form of
the two-time correlation function
\begin{eqnarray}
\lim_{t\rightarrow\infty}\left\langle \sigma_{-}(t+\tau)\sigma_{+}(t)\right\rangle  & = & \frac{1}{4}\sum_{n\,{\rm odd}}\mathrm{Tr}_{SR}[\tilde{U}^{\dagger}(\tau)\hat{C}_{n}^{\dagger}\tilde{U}(\tau)\hat{C}_{n}\tilde{\rho}(\infty)\rho_{R}]e^{-in\omega\tau}\nonumber \\
 & = & \frac{1}{4}\sum_{n\,{\rm odd}}\mathrm{Tr}_{S}\{\hat{C}_{n}^{\dagger}\mathrm{Tr}_{R}[\tilde{U}(\tau)\hat{C}_{n}\tilde{\rho}(\infty)\rho_{R}\tilde{U}^{\dagger}(\tau)]\}e^{-in\omega\tau}\nonumber \\
 & = & \frac{1}{4}\sum_{n\,{\rm odd}}\mathrm{Tr}_{S}e^{-in\omega\tau}[\hat{C}_{n}^{\dagger}\tilde{\rho}(\tau)]|_{\tilde{\rho}(0)=\hat{C}_n\tilde{\rho}(\infty)}\nonumber\\
&\equiv&\frac{1}{4}\sum_{n\,{\rm odd}}e^{-in\omega\tau}\langle\hat{C}^{\dagger}_{n}(\tau)\rangle|_{\tilde{\rho}(0)=\hat{C}_n\tilde{\rho}(\infty)},\label{eq:taufun1}
\end{eqnarray}
where $\tilde{\rho}(\tau)$ is the solution of Eq.~(\ref{eq:menew})
with initial condition $\tilde{\rho}(0)=\hat{C}_{n}\tilde{\rho}(\infty)$. When deriving the third line in Eq.~(\ref{eq:taufun1}), we used quantum regression theory~\cite{Lax}. Similarly, the other two-time correlation function takes the same form as the last line in Eq.~(\ref{eq:taufun1}) but with initial condition $\tilde{\rho}(0)=\tilde{\rho}(\infty)\hat{C}_{n}$.

Finally, we find that the two-time commutator is connected with the quantities in the transformed frame as follows:
\begin{eqnarray}
\lim_{t\rightarrow\infty}\left\langle [\sigma_{-}(t+\tau),\sigma_{+}(t)]\right\rangle  & = & \frac{1}{4}\sum_{n\,{\rm odd}}e^{-in\omega\tau}[\langle\hat{C}_{n}^{\dagger}(\tau)\rangle|_{\tilde{\rho}(0)=\hat{C}_{n}\tilde{\rho}(\infty)}-\langle\hat{C}_{n}^{\dagger}(\tau)\rangle|_{\tilde{\rho}(0)=\tilde{\rho}(\infty)\hat{C}_{n}}]\nonumber \\
 & = & \frac{1}{4}\sum_{n\,{\rm odd}}e^{-in\omega\tau}\langle\langle\hat{C}_{n}^{\dagger}(\tau)\rangle\rangle|_{\tilde{\rho}(0)=\hat{C}_{n}\tilde{\rho}(\infty)-\tilde{\rho}(\infty)\hat{C}_{n}},
\end{eqnarray}
where $\langle\langle\hat{C}_n^\dagger(\tau)\rangle\rangle$ denotes mean value of $\hat{C}^\dagger_n$ averaged over the solution to the homogeneous part of master equation (\ref{eq:menew}) with initial condition $\tilde{\rho}(0)=\hat{C}_{n}\tilde{\rho}(\infty)-\tilde{\rho}(\infty)\hat{C}_{n}$.
Substituting this expression into Eq.~(\ref{eq:sfun}) and letting $p=-i(\upsilon-n\omega)$, we can express the spectrum function in terms of the Laplace transform of $\langle\langle\hat{C}_{n}^{\dagger}(\tau)\rangle\rangle$. Since $\langle\langle\hat{C}_{n}^{\dagger}(\tau)\rangle\rangle=f_{+,n}^{+}\langle\langle \tilde{s}_{-}(\tau)\rangle\rangle+f_{-,n}^{+}\langle\langle \tilde{s}_{+}(\tau)\rangle\rangle+f_{z,n}^{+}\langle\langle \tilde{s}_{z}(\tau)\rangle\rangle$, the spectrum function can be expressed as follows:
\begin{equation}
S(\upsilon)\propto\frac{1}{4}\mathrm{Re}\sum_{n\,{\rm odd}}[f_{+,n}^{+}g_{-}(p)+f_{-,n}^{+}g_{+}(p)+f_{z,n}^{+}g(p)]|_{p=-i(\upsilon-n\omega)},
\end{equation}
where $g_j(p)$ ($j=\pm, z$) are given in Eqs.~(\ref{eq:sxlp})-(\ref{eq:szlp}) and their initial conditions are given by
\begin{eqnarray}
x_{0} & = & \mathrm{Tr}\{[s_{+},\hat{C}_{n}]\tilde{\rho}(\infty)\}\nonumber \\
 & = & f_{-,n}^{+}\langle\tilde{s}_{z}\rangle_{\mathrm{ss}}-2f_{z,n}^{+}\langle\tilde{s}_{+}\rangle_{\mathrm{ss}},\\
y_{0} & = & \mathrm{Tr}\{[s_{-},\hat{C}_{n}]\tilde{\rho}(\infty)\}\nonumber \\
 & = & -f_{+,n}^{+}\langle\tilde{s}_{z}\rangle_{\mathrm{ss}}+2f_{z,n}^{+}\langle\tilde{s}_{-}\rangle_{\mathrm{ss}},\\
z_{0} & = & \mathrm{Tr}\{[s_{z},\hat{C}_{n}]\tilde{\rho}(\infty)\}\nonumber \\
 & = & 2f_{+,n}^{+}\langle\tilde{s}_{+}\rangle_{\mathrm{ss}}-2f_{-,n}^{+}\langle\tilde{s}_{-}\rangle_{\mathrm{ss}}.
\end{eqnarray}

\begin{figure}
  \includegraphics[width=8cm]{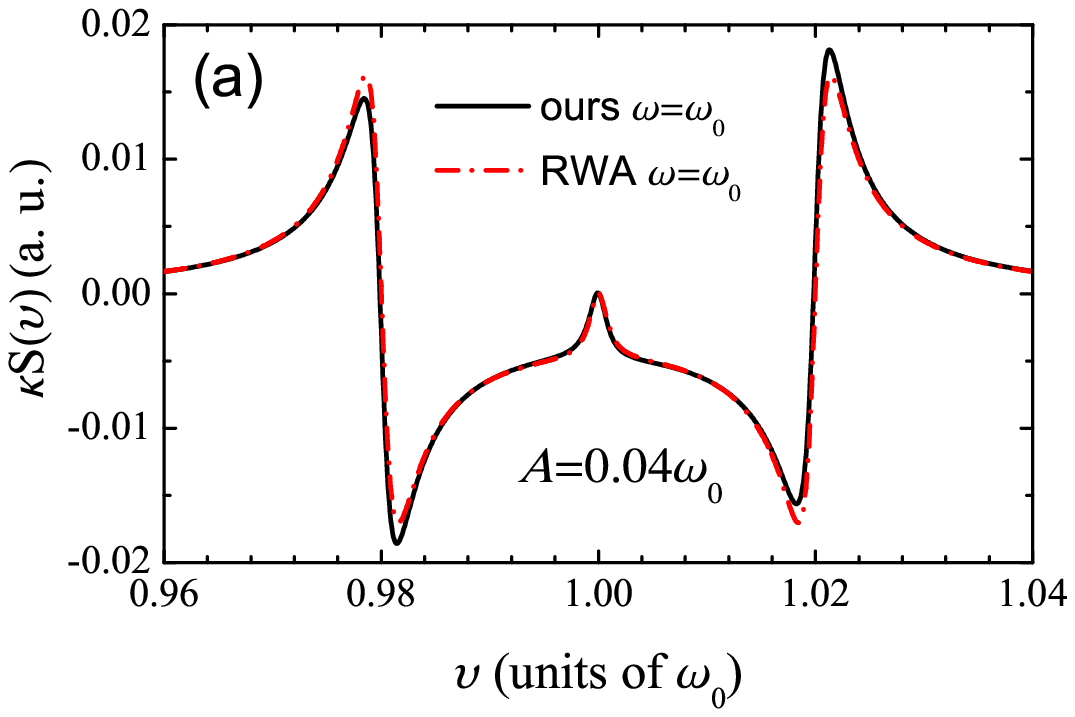}
  \includegraphics[width=8cm]{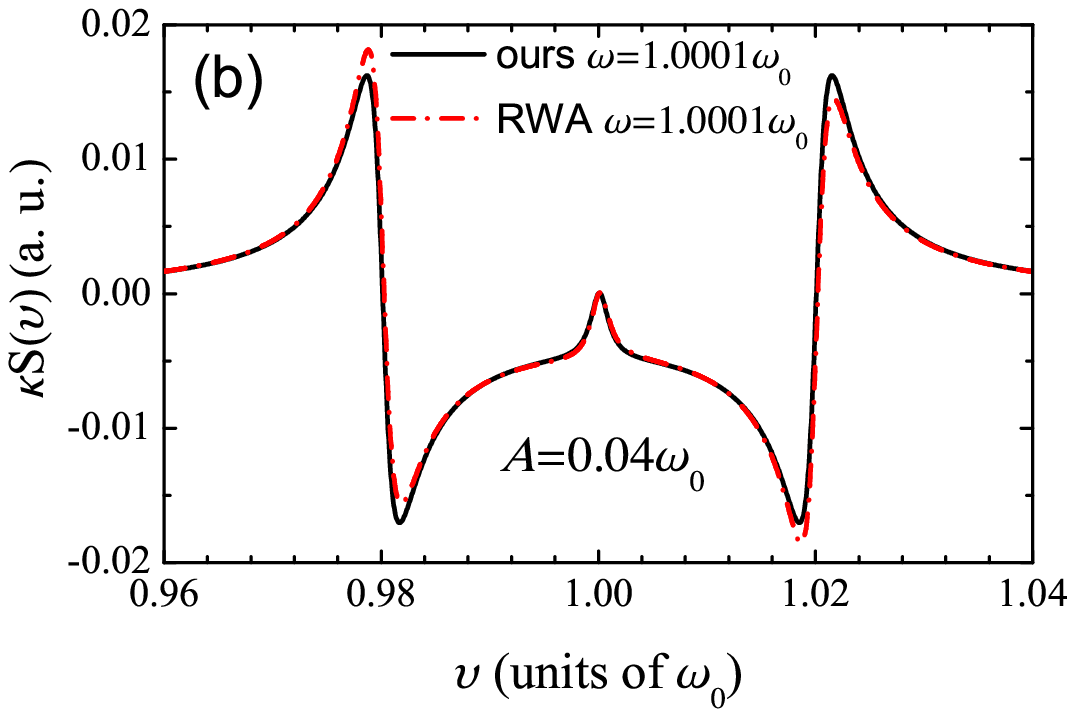}
  \includegraphics[width=8cm]{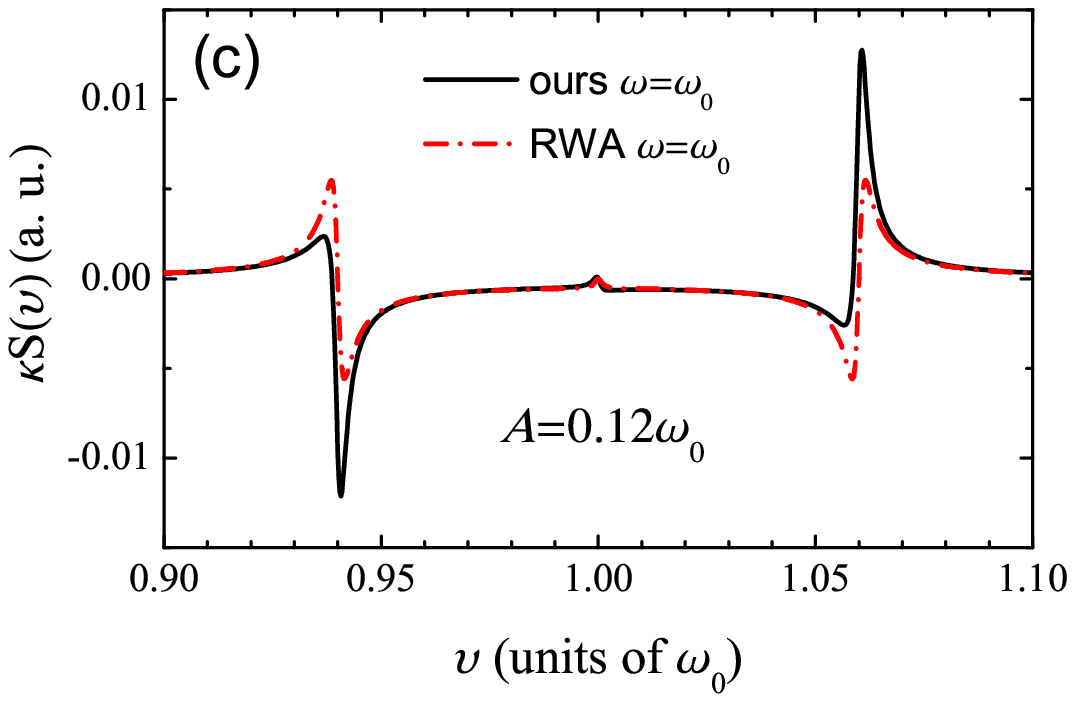}
  \includegraphics[width=8cm]{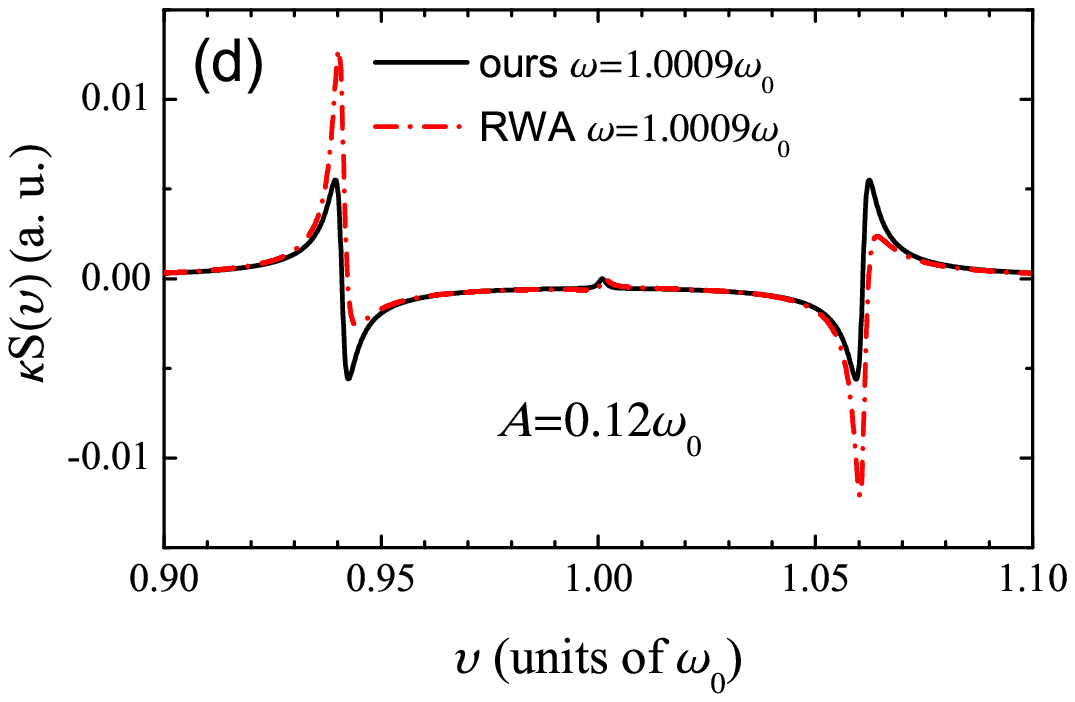}
  \caption{(Color online) The probe-pump spectrum $S(\upsilon)$ is shown as a function of probe frequency $\upsilon$ for $\kappa=2\times10^{-3}\omega_0$.}\label{fig4}
\end{figure}

Now we illustrate the role of BS shift in the probe-pump spectrum calculated without the RWA. In Figs.~\ref{fig4}(a) and~\ref{fig4}(c), we show the comparison between the non-RWA and RWA spectra for a fixed driving strength under the resonant condition $\omega=\omega_0$ of the RWA. We notice that the RWA spectra are always symmetric with respect to the line $\omega=\omega_0$ while the non-RWA spectra are asymmetric and distinguish clearly from the RWA cases when $A$ increases. In this case, the lineshapes for the non-RWA resemble those of the RWA at a detuning case [see Figs.~\ref{fig4}(b) and~\ref{fig4}(d)]. In fact, this asymmetric behavior results from the BS shift. In Figs.~\ref{fig4}(b) and~\ref{fig4}(d), we show that symmetric non-RWA spectra appear for $\omega=\omega_0+\delta\omega_{\rm BS}\equiv\omega_{\mathrm{res}}$ in comparison with the RWA asymmetric lineshapes. These results indicate that when the BS shift is correctly compensated for driving frequency, the probe-pump spectra become symmetric and is similar as the RWA ones for $\omega=\omega_0$.

\begin{figure}
  \includegraphics[width=8cm]{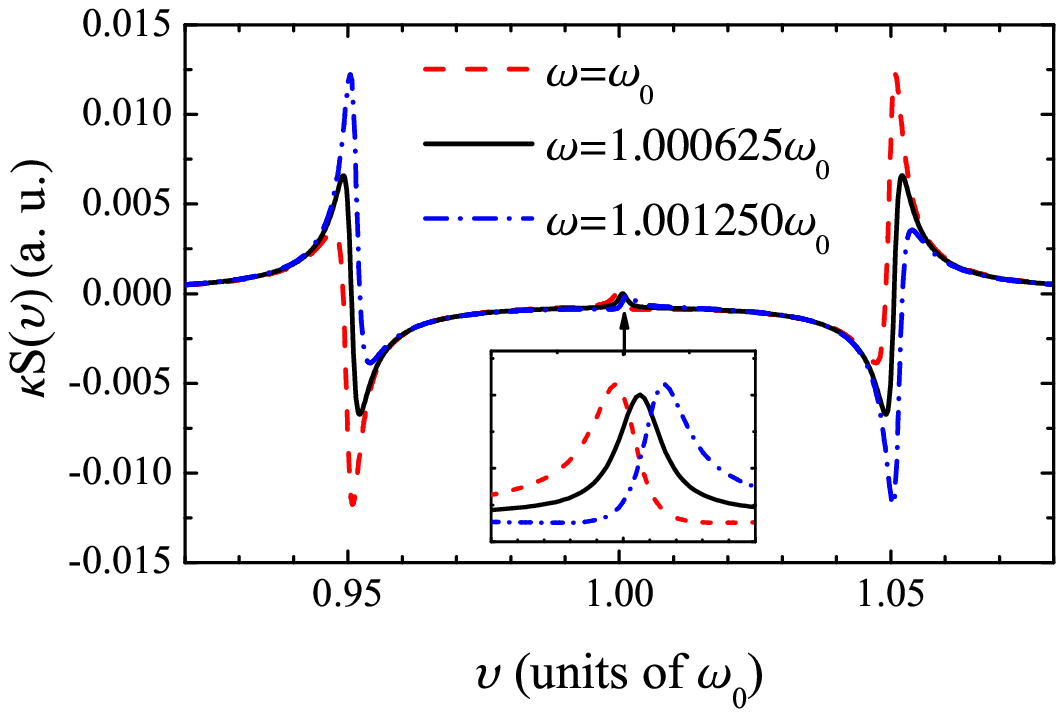}
  \caption{(Color online) The probe-pump spectrum $S(\upsilon)$ is shown as a function of probe frequency $\upsilon$ for $\kappa=2\times10^{-3}\omega_0$ with $A=0.1\omega_0$.}\label{fig5}
\end{figure}

In Fig.~\ref{fig5}, we show the non-RWA spectra for $A=0.1\omega_0$ with three different pump frequencies. Although the detunings of the pump field are very small, one can observe difference in the lineshapes of the spectra. We notice that for $\omega= \omega_{\mathrm{res}}=\omega_0+\delta\omega_{\rm BS} ~(\delta\omega_{\rm BS}=0.000625\omega_0 )$, the curve exhibits two equally symmetric sidebands in the wing. However, for $\omega \neq\omega_{\mathrm{res}}$, the lineshapes are clearly asymmetric (the spectra of $\omega=\omega_{\rm res}\pm\delta\omega_{\rm BS}$ are shown in the red-dashed line and the blue dotted-dashed line, respectively). Therefore, the probe-pump spectrum provides a way to sense the roles of the BS shift by two properties: (i) for $\omega=\omega_0$, the lineshape of the spectrum changes from nearly symmetric to asymmetric as $A$ increases, while that of the RWA are always symmetric; (ii) the non-RWA lineshape becomes symmetric only under the exact-resonance condition, i.e. $\omega=\omega_0+\delta\omega_{\mathrm{BS}}$. One expects that the two properties can be checked in superconducting-circuit qubits under the moderately strong driving field.

\section{Conclusion}
\underline{}In summary, we have calculated the BS shift over the entire driving-strength range using the derivative of the effective Rabi frequency and demonstrated detectable signatures induced by the BS shift. It turned out that our method can be correctly applied to study the small shift case when the driving strength is moderately weak, but also accurately solve the large shift case when the driving strength is sufficiently strong, which is beyond the perturbation theory. Moreover, our method allows us to examine the role of the BS shift in the emission and absorption processes from the open driven TLS. We showed that it is easy to obtain the time evolution of excited-state population of the TLS as well as its steady-state behavior, which indicates the emergence of resonance. We illustrated that the time-averaged population of TLS provides a direct measurement of the BS shift, which is consistent with previous work. Furthermore, we found that in experiment accessible parameters regimes, the non-RWA spectrum becomes symmetric only for $\omega=\omega_0+\delta\omega_{\rm BS}$, i.e. for the resonance condition correctly taken into account the BS shift. Otherwise, the lineshape of the spectrum is generally asymmetric. While for $\omega=\omega_0$, the non-RWA spectrum becomes asymmetric with the increase of the driving strength, while the RWA spectrum is always symmetric.

\begin{acknowledgments}
  This work was supported by the National Natural Science Foundation of China (Grants No.~11174198, No.~11374208, No.~91221201, and No.~11474200) and the National Basic Research Program of China (Grant No.~2011CB922202). The work was partially supported by the Shanghai Jiao Tong University SMC-Youth Foundation.
\end{acknowledgments}
\appendix

\section{The derivation of master equation in the transformed frame\label{sub:The-derivation}}
We show the derivation of the master equation [Eq.~(\ref{eq:menew})]. After the unitary transformation and rotating transformation, we obtain the transformed equation as  follows,
\begin{eqnarray}
  \frac{d}{dt}\tilde{\rho}(t)&=&-i[\tilde{H},\tilde{\rho}(t)]-\frac{\kappa}{2}[\tilde{\sigma}_{+}(t)\tilde{\sigma}_{-}(t)\tilde{\rho}(t)\nonumber\\
  & &+\tilde{\rho}(t)\tilde{\sigma}_{+}(t)\tilde{\sigma}_{-}(t)-2\tilde{\sigma}_{-}(t)\tilde{\rho}(t)\tilde{\sigma}_{+}(t)],
\end{eqnarray}
where $\tilde{\sigma}_{\pm}(t)=R(t)e^{S(t)}\sigma_\pm e^{-S(t)}R^\dagger(t)$. Employing the eigenstates of $\tilde{H}$ as the basis, we can obtain the equation of the elements of the density matrix
\begin{eqnarray}
  \frac{d}{dt}\tilde{\rho}_{\alpha\beta}(t)&=&-i(\varepsilon_\alpha-\varepsilon_\beta)\tilde{\rho}_{\alpha\beta}(t)-\frac{\kappa}{2}\langle\widetilde{\alpha}|[\tilde{\sigma}_{+}(t)\tilde{\sigma}_{-}(t)\tilde{\rho}(t)\nonumber\\
  & &+\tilde{\rho}(t)\tilde{\sigma}_{+}(t)\tilde{\sigma}_{-}(t)-2\tilde{\sigma}_{-}(t)\tilde{\rho}(t)\tilde{\sigma}_{+}(t)]|\widetilde{\beta}\rangle.
\end{eqnarray}
To obtain the explicit form of the dissipation part of the equation,  we insert the identity matrix such as $\sum_\mu |\widetilde{\nu}\rangle\langle\widetilde{\nu}|=1$ between the operators, which leads to
\begin{eqnarray}
  \langle\tilde{\alpha}|\tilde{\sigma}_{+}(t)\tilde{\sigma}_{-}(t)\tilde{\rho}(t)|\tilde{\beta}\rangle&=&\sum_{\mu,\nu}\langle\tilde{\alpha}|\tilde{\sigma}_{+}(t)|\tilde{\nu}\rangle\langle\tilde{\nu}|\tilde{\sigma}_{-}(t)|\tilde{\mu}\rangle\langle\tilde{\mu}|\tilde{\rho}(t)|\tilde{\beta}\rangle\nonumber\\
  &=&\sum_{\mu,\nu}\sum_{n,n^{\prime}}X_{\alpha\nu,n}^{+}X_{\nu\mu,n^{\prime}}^{-}e^{i(n+n^{\prime})\omega t}\tilde{\rho}_{\mu\beta}(t),\label{eq:series1}\\
  \langle\tilde{\alpha}|\tilde{\rho}(t)\tilde{\sigma}_{+}(t)\tilde{\sigma}_{-}(t)|\tilde{\beta}\rangle&=&\sum_{\mu,\nu}\sum_{n,n^{\prime}}X_{\nu\mu,n}^{+}X_{\mu\beta,n^{\prime}}^{-}e^{i(n+n^{\prime})\omega t}\tilde{\rho}_{\alpha\nu}(t),\\
  \langle\tilde{\alpha}|\tilde{\sigma}_{-}(t)\tilde{\rho}(t)\tilde{\sigma}_{+}(t)|\tilde{\beta}\rangle&=&\sum_{\mu,\nu}\sum_{n,n^{\prime}}X_{\nu\beta,n}^{+}X_{\alpha\mu,n^{\prime}}^{-}e^{i(n+n^{\prime})\omega t}\tilde{\rho}_{\mu\nu}(t),\label{eq:series3}
\end{eqnarray}
where $X_{\alpha\beta,n}^{\pm}$ are the Fourier coefficients from the expansions $\langle \widetilde{\alpha}|\tilde{\sigma}_{\pm}(t)|\widetilde{\beta}\rangle =\sum_{n}e^{in\omega t}X_{\alpha\beta,n}^{\pm}$.
To proceed, we need to give the explicit forms for $X^{\pm}_{\alpha\beta,n}$, which can be evaluated by the following integral
\begin{eqnarray}
X_{\alpha\beta,n}^{\pm} & = & \frac{\omega}{2\pi}\int_{0}^{2\pi/\omega}dt\left\langle u_{\alpha}(t)|\sigma_{\pm}|u_{\beta}(t)\right\rangle e^{-in\omega t}.
\end{eqnarray}
It is straightforward to calculate the expressions for $X_{\alpha\beta,n}^{+}$, which are given by
\begin{eqnarray}
X_{++,n}^{+} & = & \frac{1}{2}\sum_{l\:\mathrm{odd}}(f_{z,l}^{+}\delta_{l,n}+f_{z,l}^{-}\delta_{l,-n}),\\
X_{+-,n}^{+} & = & \frac{1}{2}\sum_{l\:\mathrm{odd}}(f_{+,l}^{+}\delta_{l,n}+f_{-,l}^{-}\delta_{l,-n}),\\
X_{-+,n}^{+} & = & \frac{1}{2}\sum_{l\:\mathrm{odd}}(f_{-,l}^{+}\delta_{l,n}+f_{+,l}^{-}\delta_{l,-n}),\\
X_{--,n}^{+} & = & -\frac{1}{2}\sum_{l\:\mathrm{odd}}(f_{z,l}^{+}\delta_{l,n}+f_{z,l}^{-}\delta_{l,-n}),
\end{eqnarray}
where the summation is taken over all positive odd integers, and
\begin{eqnarray}
f_{+,l}^{\pm} & = & -\left[\delta_{l,1}\pm J_{l-1}\left(\frac{A}{\omega}\xi\right)\right]\cos^{2}\theta\mp J_{l+1}\left(\frac{A}{\omega}\xi\right)\sin^{2}\theta\mp J_{l}\left(\frac{A}{\omega}\xi\right)\sin(2\theta),\label{eq:fxl}\\
f_{-,l}^{\pm} & = & \left[\delta_{l,1}\pm J_{l-1}\left(\frac{A}{\omega}\xi\right)\right]\sin^{2}\theta\pm J_{l+1}\left(\frac{A}{\omega}\xi\right)\cos^{2}\theta\mp J_{l}\left(\frac{A}{\omega}\xi\right)\sin(2\theta),\\
f_{z,l}^{\pm} & = & \frac{1}{2}\left[\delta_{l,1}\pm J_{l-1}\left(\frac{A}{\omega}\xi\right)\mp J_{l+1}\left(\frac{A}{\omega}\xi\right)\right]\sin(2\theta)\mp J_{l}\left(\frac{A}{\omega}\xi\right)\cos(2\theta).\label{eq:fzl}
\end{eqnarray}
The explicit expressions for $X_{\alpha\beta,n}^{-}$ can be directly obtained
by the relation
\begin{equation}
X_{\alpha\beta,n}^{-}=(X_{\beta\alpha,-n}^{+})^{\ast}.
\end{equation}
When calculating $X^+_{\alpha\beta,n}$, we used the identities
\begin{equation}
R(t)e^{S(t)}\sigma_{+}e^{-S(t)}R^{\dagger}(t)=\frac{1}{2}\sum_{n\,\mathrm{odd}}(\hat{C}_{n}e^{in\omega t}+\hat{D}_{n}e^{-in\omega t}),\label{eq:identity}
\end{equation}
where
\begin{eqnarray}
\hat{C}_{n} & = & f_{+,n}^{+}s_{+}+f_{-,n}^{+}s_{-}+f_{z,n}^{+}s_{z},\label{eq:cn}\\
\hat{D}_{n} & = & f_{-,n}^{-}s_{+}+f_{+,n}^{-}s_{-}+f_{z,n}^{-}s_{z}.\label{eq:dn}
\end{eqnarray}
Here, we introduced a set of dressed-state operators:
\begin{equation}
s_{+} = s_{-}^{\dagger}=|\widetilde{+}\rangle\langle\widetilde{-}|,\quad
s_{z}  =  |\widetilde{+}\rangle\langle\widetilde{+}|-|\widetilde{-}\rangle\langle\widetilde{-}|.\label{eq:szde}
\end{equation}

It turns out that the master equation in the new representation still possesses explicit time-dependence. However, in the strong-driving regime, it is feasible to remove the explicit time-dependence by invoking the partial secular approximation (Moderate RWA)~\cite{Grifoni,Grifoni2}. In other words, we only keep the terms satisfying $n+n^\prime=0$ in Eqs.~(\ref{eq:series1})-(\ref{eq:series3}). This approximation can be justified when $\tilde{\Omega}_R\gg \kappa$. Consequently, we arrive at the time-independent master equation in Eq.~(\ref{eq:menew}). One can verify that the time-independent master equation (\ref{eq:menew}) can predict almost the same dynamics as that given by Eq.~(\ref{eq:me}) when $\tilde{\Omega}_R\gg \kappa$.

\section{The solutions to the master equation in the transformed frame}\label{app:sols}

The master equation can be rewritten in terms of the mean value of dressed-state
operators by the following relations:
\begin{equation}
  \langle \tilde{s}_z(t)\rangle=\tilde{\rho}_{++}(t)-\tilde{\rho}_{--}(t),\quad\langle \tilde{s}_+(t)\rangle=\langle \tilde{s}_-(t)\rangle^\ast=\tilde{\rho}_{-+}(t),
\end{equation}
\begin{eqnarray}
\frac{d}{dt}\left\langle \tilde{s}_{z}(t)\right\rangle  & = & -\gamma_{z}\left\langle \tilde{s}_{z}(t)\right\rangle -2\gamma_{1}(\left\langle \tilde{s}_{+}(t)\right\rangle +\left\langle \tilde{s}_{-}(t)\right\rangle )-\gamma_{0},\label{eq:szt}\\
\frac{d}{dt}\left\langle \tilde{s}_{+}(t)\right\rangle  & = & \frac{d}{dt}\left\langle \tilde{s}_{-}(t)\right\rangle ^{\ast}\nonumber \\
 & = & i\tilde{\Omega}_{R}\left\langle \tilde{s}_{+}(t)\right\rangle -\gamma_{1}\left\langle \tilde{s}_{z}(t)\right\rangle -\gamma_{-}\left\langle \tilde{s}_{-}(t)\right\rangle -\gamma_{+}\left\langle \tilde{s}_{+}(t)\right\rangle -\gamma_{2},\label{eq:sxt}
\end{eqnarray}
where
\begin{eqnarray}
\gamma_{z} & = & {\cal L}_{++,++}-{\cal L}_{++,--},\nonumber \\
\gamma_{0} & = & {\cal L}_{++,++}+{\cal L}_{++,--},\nonumber \\
\gamma_{1} & = & {\cal L}_{++,+-},\nonumber \\
\gamma_{2} & = & ({\cal L}_{-+,++}+{\cal L}_{-+,--})/2,\nonumber \\
\gamma_{-} & = & {\cal L}_{-+,+-},\nonumber \\
\gamma_{+} & = & {\cal L}_{-+,-+}.
\end{eqnarray}
On solving the equations, we can fully determine the evolution of the TLS. In particular, the steady-state solutions can be easily found as follows:
\begin{eqnarray}
\left\langle \tilde{s}_{z}\right\rangle _{\mathrm{ss}} & = & \frac{-\tilde{\Omega}_{R}^{2}\gamma_{0}-4\gamma_{1}\gamma_{2}(\gamma_{-}-\gamma_{+})+\gamma_{0}(\gamma_{-}^{2}-\gamma_{+}^{2})}{4\gamma_{1}^{2}(\gamma_{-}-\gamma_{+})+(\tilde{\Omega}_{R}^{2}-\gamma_{-}^{2}+\gamma_{+}^{2})\gamma_{z}},\label{eq:szss}\\
\left\langle \tilde{s}_{+}\right\rangle _{\mathrm{ss}} & = & \left\langle \tilde{s}_{-}\right\rangle _{\mathrm{ss}}^{\ast}=\frac{(i\tilde{\Omega}_{R}-\gamma_{-}+\gamma_{+})(\gamma_{0}\gamma_{1}-\gamma_{2}\gamma_{z})}{4\gamma_{1}^{2}(\gamma_{-}-\gamma_{+})+(\tilde{\Omega}_{R}^{2}-\gamma_{-}^{2}+\gamma_{+}^{2})\gamma_{z}}.\label{eq:cohss}
\end{eqnarray}

When calculating the probe-pump spectrum, we need the solutions to homogeneous parts of Eqs.~(\ref{eq:menew}). The corresponding homogeneous differential equations can be solved by Laplace transform. Denoting the solutions as $\langle\langle\tilde{s}_j(t)\rangle\rangle$ ($j=\pm, z$), we can obtain their Laplace transforms,
\begin{eqnarray}
g_{+}(p) & = & \int^\infty_0 e^{-p t}\langle\langle\tilde{s}_+(t)\rangle\rangle dt\nonumber\\
& = &\frac{1}{F(p)}\{x_{0}[(p+\gamma_{+}+i\tilde{\Omega}_{R})(p+\gamma_{z})-2\gamma_{1}^{2}]+y_{0}[2\gamma_{1}^{2}-\gamma_{-}(p+\gamma_{z})]\nonumber \\
 &  & -\gamma_{1}z_{0}(p+i\tilde{\Omega}_{R}-\gamma_{-}+\gamma_{+})\},\label{eq:sxlp}\\
g_{-}(p) & = & \int^\infty_0 e^{-p t}\langle\langle\tilde{s}_-(t)\rangle\rangle dt\nonumber\\
& = & \frac{1}{F(p)}\{y_{0}[(p+\gamma_{+}-i\tilde{\Omega}_{R})(p+\gamma_{z})-2\gamma_{1}^{2}]+x_{0}[2\gamma_{1}^{2}-\gamma_{-}(p+\gamma_{z})]\nonumber \\
 &  & -\gamma_{1}z_{0}(p-i\tilde{\Omega}_{R}-\gamma_{-}+\gamma_{+})\},\\
g_{z}(p) & = & \int^\infty_0 e^{-p t}\langle\langle\tilde{s}_z(t)\rangle\rangle dt\nonumber\\& = & \frac{1}{F(p)}\{z_{0}[(p+\gamma_{+})^{2}+\tilde{\Omega}_{R}^{2}-\gamma_{-}^{2}]-2\gamma_{1}x_{0}(p+i\tilde{\Omega}_{R}-\gamma_{-}+\gamma_{+})\nonumber \\
 &  & -2\gamma_{1}y_{0}(p-i\tilde{\Omega}_{R}-\gamma_{-}+\gamma_{+})\},\label{eq:szlp}
\end{eqnarray}
where the initial conditions are $x_0=\mathrm{Tr}[s_+\tilde{\rho}(0)]$, $y_0=\mathrm{Tr}[s_-\tilde{\rho}(0)]$ and $z_0=\mathrm{Tr}[s_z\tilde{\rho}(0)]$, and the polynomial $F(p)$ is given by
\begin{eqnarray}
F(p) & = & p^{3}+4\gamma_{1}^{2}(\gamma_{-}-\gamma_{+})+(\tilde{\Omega}_{R}^{2}-\gamma_{-}^{2}+\gamma_{+}^{2})\gamma_{z}\nonumber \\
 &  & +p^{2}(\gamma_{z}+2\gamma_{+})+p(\tilde{\Omega}_{R}^{2}-4\gamma_{1}^{2}-\gamma_{-}^{2}+\gamma_{+}^{2}+2\gamma_{+}\gamma_{z}).
\end{eqnarray}

\begin{table}
\begin{tabular}{ccccc}
\hline\hline
$A/\omega_{0}$ & numerical & present paper & Eq.~(\ref{q2}) & Eq.~(\ref{eq:asym})\\
\hline
1.0 & 0.063224 & 0.063268 & 0.063228 & \\

3.5 & 0.707959 & 0.716200 & 0.712320 & 0.455407\\

6.0 & 1.641809 & 1.649924 & 1.650482 & 1.494983\\

8.5 & 2.637787 & 2.640075 & 2.639255 & 2.534559\\

11.0 & 3.653740 & 3.652351 & 3.641373 & 3.574136\\

13.5 & 4.678502 & 4.675271 & 4.650384 & 4.613712\\

16.0 & 5.707919 & 5.703825 & 5.664602 & 5.653289\\

18.5 & 6.740093 & 6.735637 & 6.683190 & 6.692864\\

21.0 & 7.774035 & 7.769474 & 7.705492 & 7.732441\\
\hline
\hline
\end{tabular}

\caption{The comparison of the BS shift obtained by various methods.}\label{tab:table1}
\end{table}

\end{document}